\documentclass{article}
\usepackage{amsthm}
\usepackage{epsf}
\usepackage{latexsym}
\usepackage{amsmath}
\usepackage{amsfonts}
\usepackage{amssymb}
\usepackage{graphicx,epsf}
\usepackage{dsfont}
\usepackage[]{latexsym}

\newcommand{\be}{\begin{equation}}
\newcommand{\ee}{\end{equation}}
\newcommand{\ba}{\begin{eqnarray}}
\newcommand{\ea}{\end{eqnarray}}

\begin{document}
\date{}
\title{On the gauge features of gravity on a Lie algebroid structure}

\author{S. Fabi\thanks{Email: fabi001@bama.ua.edu}~, B. Harms\thanks{Email: bharms@bama.ua.edu}~ and S. Hou\thanks{Email: shou@crimson.ua.edu}\\
\\
\emph{Department of Physics and Astronomy}\\
 \emph{The University of Alabama}\\
\emph{ Box 870324, Tuscaloosa, AL 35487-0324, USA}}

\maketitle

\begin{abstract}
    {We present the geometric formulation of gravity based on
    the mathematical structure of a Lie Algebroid. We show that this framework provides the geometrical setting to describe the gauge propriety of gravity.}
\end{abstract}

\tableofcontents

\section{Introduction}
Understanding the geometrical interpretation of gravity
has been a goal since Einstein formulated General Relativity (GR) in 1915.
In Einstein's theory the effects of gravity are described by the metric tensor $g_{\mu\nu}(x)$
from which the Riemann tensor, via the Levi-Civita connection, can be calculated,
and gravity is understood as the measurement of the curvature of spacetime.

General relativity is not the only theory based on a geometrical
interpretation of gravity. Other theories both, within and without the
context of Riemann geometry exist, and they describe the
gravitational interaction in terms of different geometrical
interpretations. Examples are: teleparallel gravity, Einstein-Cartan
theory and metric-affine gravity \cite{hehl1995}. Even outside
the conventional scheme of a smooth manifold, there have been
several attempts to maintain a geometric meaning for spacetime
 as in the case of noncommutative geometry or spin networks; but at the quantum gravity level,
the geometric structure of spacetime might turn out to be so complex
that people have suggested giving up a possible geometric
interpretation of gravity \cite {Weinberg:1972}.

Given the considerations mentioned above, we choose to study further the link
between gravity and geometry. The fundamental assumption is that the
torsion $T(x)^c_{~ab}$, can be consider to be the geometrical field
which characterizes a spacetime in which gravity is present, as in
teleparallel gravity \cite{pereira2010}. This approach leads us to a
new gauge interpretation of the theory since now the torsion's
coefficient represents the structure ``constants'' of the
generalized momentum generators $D_a$ \cite{hehl1995} such that
$[D(x)_a,D(x)_b]=-T(x)^{c}_{~ab}D_c$ which was introduced in
\cite{Fabi:2011mk}. This paper presents therefore a geometrical
interpretation of gravity and especially how its gauge properties
\cite {Trautman:1970cy, Ivanenko:1984vf, Guay:2004zz} emerge naturally
on the geometrical structure of a Lie algebroid. We call this Lie
algebroid the gravity Lie algebroid.

 The use of an algebroid to describe gravity has also been considered in the context of string theory
  where the Lie algebroid is the co-tangent space of a manifold together with a twisted Koszul bracket \cite{Blumenhagen:2012nt}.
 In \cite{Blohmann:2010jd} the groupoid of the diffeomorphisms of space-like hypersurfaces in spacetimes is considered as a description of gravity
 (as general relativity in the constrained hamiltonian formulation).


\section{Gravity and gauge theories}

Yang-Mills theory is a gauge theory with respect to the internal symmetries and
the principle bundles provide the geometrical setting to express its classical features
\cite{Trautman:1970cy, Guay:2004zz, Daniel:1979ez, Catren:2008zz}.
 The fiber, or structure group, is $U(1)$ or $SU(2)$ or $SU(3)$ to describe respectively electromagnetism, weak, or strong interactions.
 The aim of a gauge theory of gravity is to extend this description to gravity. This research project was started by Utiyama \cite{Utiyama1956}, but both the nature of the gauge group for gravity and the geometrical setting of the theory still present difficulties \cite{Tresguerres2002}.

Utiyama's work introduced as gauge potentials (or compensating fields as the gauge prescription requires)
 a set of fields with the geometrical interpretation of spin connections. In order to obtain an invariant matter Lagrangian,
he had to strategically add an extra set of compensating fields: the tetrads $h_a^{~\mu}$ which did not arise as gauge potentials.
Later Kibble \cite{Kibble1961,Blago2002}, proposed the Poincare group as the gauge group, and the tetrads appeared naturally in the theory
as functions of the gauge potentials of the translational part of the Poincare group. Using both the spin connection $\omega_{\mu}$ and the tetrad $h^a_{~\mu}$, the
`Poincare' covariant derivative turns out to be
\be\label{pocov}
\nabla_a=h_a^{~\mu}\nabla_{\mu}.
\ee
Where $\nabla_{\mu}=\partial_{\mu}+\omega_{\mu}$ and $h_a^{~\mu}=\delta_a^{~\mu}+B_a^{~\mu}$ with the $B_a^{~\mu}$ taken as the translation part
of the connection (or translation gauge potentials).
 However localizing the Poincare group (Sciama-Kibble theory)
 does not reproduce GR but gives rise to the Einstein-Cartan theory, in which the
energy is the source of curvature, and the spin density is the source of torsion \cite{DeSabbata1986}.
In \cite{Kleinert1998} Kleinert criticizes this interpretation of the torsion since it relates the torsion to only
the spin context of the theory and not to the total angular momentum $J=L+S$. One physical example, which
shows how this interpretation of the torsion is problematic, is
the decay of the spin-1 meson $\rho$ into the two spin-0 pions $\pi$. This decay is explained by
taking $J$ as the physical degree of freedom which is conserved, since $L$ or $S$ are not independently conserved.
During the same period of the pioneer works by Utiyama, M{\o}ller \cite{Moller1961} reconsidered the theory of teleparallel gravity (originally
proposed by Einstein in 1928 as a model of unification of GR and electromagnetism), and in 1979 Hayashi and Shirafuji \cite{Hayashi1979}
showed that teleparallel gravity is a gauge theory of the translation group, where the corresponding field strength is given
by the torsion and is the only field (not the curvature) responsible for the gravitational interaction.
They also showed that one of the particular expressions of teleparallel gravity becomes equivalent to GR
(and in this case the theory is referred to as the teleparallel equivalent of GR).
In the work by Hehl \cite{hehl1995}, \cite{Gronwald:1995em} and Blagojevic \cite{Blago2002} teleparallel gravity is considered
as a particular case of the Sciama-Kibble theory corresponding to zero curvature.
 Pereira \cite{pereira2010} points out a few issues with the phenomenology of their approach due to the problem of
a non-propagating torsion field present in their theory.

A different approach associated with localizing the Poincare group \cite{Banados:1996hi,Regge:1986nz}, is based on a connection of the form
\be\label{pocon}
A_{\mu}=h^a_{~\mu}p_a+A^{ab}_{~~\mu}J_{ab},
\ee
where the $p_a$ and the $J_{ab}$ are the translation and angular momentum generators respectively.
When considering the free Lagrangian constructed from this gauge potential in first order formalism (i.e the Lagrangian for the gravitation field with
 variables given by the tetrads and spin connections), the corresponding action is not gauge invariant under translations.
  An additional problem with (\ref{pocon}) consists in taking the tetrads as the gauge potential \cite{Ivanenko:1984vf}.
In fact in the fiber bundle setting the tetrad bases are sections and not connections,
and they transform as covectors \cite{Tresguerres2002}.
Also different bundle structures have been proposed for the Poincare gauge theory, as for example the composite fiber bundle \cite{Tresguerres2002}.
These problems motivated a better understanding of localizing translations with the appropriate gauge potential and phenomenology.
The gravity Lie algebroid scenario which we present in this paper offers an alternative solution to the problem of finding the
geometrical setting of a gauge theory of gravity.

\section{Teleparallel Gravity}
We begin by reviewing teleparallel gravity \footnote {We use the
term ``Teleparallel gravity '' to refer the theory originally named
as `` Teleparallel equivalent of GR '', for clarification see
\cite{pereira2010}.}. The basic aspect of the theory is the
interpretation of the torsion: it is the dynamical quantity which is
non-zero when a gravitational field is present (while the curvature
of the connection is always zero). We review briefly the basic
geometrical and gauge invariant aspects
 (for a detailed introduction and formulation see \cite{pereira2010}).

\subsection{Geometrical aspects} The tetrad postulate states that
the covariant derivative of the tetrad $h^a_{~\rho}(x)$ is zero \be
\partial_{\mu}h^a_{~\nu}-\Gamma^{\rho}_{~\mu\nu}h^a_{~\rho}+A^a_{~b\nu}h^b_{~\nu}=0,
\ee
which can be written as
\be\label{glc}
\Gamma^{\rho}_{~\nu\mu}=h_a^{~\rho}\partial_{\mu}h^a_{~\nu}+h_a^{~\rho}A^a_{~b\mu}h^b_{~\nu},
\ee
where $\Gamma^{\rho}_{~\nu\mu}$ is a general linear connection.
 The Weitzenb\"{o}ck connection is the term
 \footnote{In \cite{pereira2010} (Eq. 3.86),
 the Weitzenb\"{o}ck connection is defined with the additional term $\dot{A}^a_{~b\mu}$. If non-zero components of $\dot{A}^a_{~b\mu}$ appear,
  they are not associated with the presence of a gravitational field, but with a non-inertial frame;
  similar to the case of non-zero elements of the Levi-Civita connection in GR  when it's expressed in a curvilinear coordinate system for the Minkowski flat spacetime.
For simplicity, we set $\dot{A}^a_{~b\mu}=0$ and since gravity is fully described by the Weitzenb\"{o}ck connection as defined in (\ref{Wcon}).}
\be\label{Wcon}
\dot{\Gamma}^{\rho}_{~\nu\mu}\equiv h_a^{~\rho}\partial_{\mu}h^{a}_{~\nu},
\ee
and $A^a_{~b\nu}$ is the spin connection. The curvature of the Weitzenb\"{o}ck connection is zero.
In teleparallel gravity the spin connection vanishes, and the torsion
is defined in terms of the Weitzenb\"{o}ck connection only as
\be
T^{\rho}_{~\mu\nu}\equiv \dot{\Gamma}^{\rho}_{~\nu\mu}-\dot{\Gamma}^{\rho}_{~\mu\nu},
\ee
or in terms of the tetrad
\be
T^{a}_{~\mu\nu}=\partial_{\mu}h^a_{~\nu}-\partial_{\nu}h^a_{~\mu}.
\ee
Alternatively to equation (\ref{glc}),
a general linear connection can be expressed as
\be
\Gamma^{\rho}_{~\mu\nu}= \genfrac{\{}{\}}{0pt}{1}{\rho}{\mu\nu}+K^{\rho}_{~\mu\nu},
\ee
where the Christoffel symbols represent the Levi-Civita connection of GR and $K^{\rho}_{~\mu\nu}$ is the contortion
\be
K^{\rho}_{~\mu\nu}= \tfrac{1}{2}(T_{\mu~\nu}^{~\rho}+T_{\nu~\mu}^{~\rho}-T^{\rho}_{~\mu\nu}).
\ee

\subsection{Gauge invariant aspects} Teleparallel gravity is a valid gauge theory of the local translation group with the gauge
potential \be B_{\mu}=B^a_{~\mu}p_a \ee with values in the Lie algebra of the momentum generators $p_a=\partial_{x^a}$.
 The $x^a=x^a(x^{\mu})$ are the coordinates of the `internal' affine
space on which the gauge transformations of local translations take
place\footnote{The coordinates $x^a$ correspond to the translational Goldstone fields $\xi^a$
  in the geometrical setting of \cite{Tresguerres2008}}.
\be\label{ltra}
x^a\rightarrow x^a+\epsilon^a (x).
\ee
This implies for a general field $\psi(x^{\mu},x^a)$ a variation
\be
\delta\psi=-\epsilon^a\partial_a\psi,
\ee
and for the gauge potential $B^a_{~\mu}$
\be
\delta B^a_{~\mu}=\partial_{\mu}\epsilon^a.
\ee
The translational covariant derivative is
\be\label{covder}
D_{\mu}=\partial_{\mu}+B_{\mu}=h^a_{~\mu}\partial_{a},
\ee
with \footnote{ We use $\delta^{a}_{~\mu}$ instead of $\partial_{\mu}x^a$ since the non-holomicity is carried by
the gauge potential and follow \cite{Gronwald:1995em,Blago2002}.}
\be
h^a_{~\mu}=\delta^a_{~\mu}+B^a_{~\mu}\equiv D_{\mu}x^a.
\ee

The torsion is related to the gauge potential via the covariant derivatives (in this sense
it's the \emph{curvature} of the \emph{connection} $B^a_{~\mu}$ )
\be
T^{a}_{~\mu\nu}=\partial_{\mu}B^a_{~\nu}-\partial_{\nu} B^a_{~\mu},
\ee
and is the field strength obtained from the commutator of the covariant derivatives
\be
[D_{\mu},D_{\nu}]\psi=T_{\mu\nu}\psi
\ee
with
\be
T_{\mu\nu}= T^a_{~\mu\nu}\partial_{a}.
\ee
This shows that the torsion is an element of the algebra of translations on $x^a$.
The Lagrangian is quadratic in the field strength, the same as the Lagrangian for Yang-Mills theory
\be
\mathcal{L}=\tfrac{h}{4k^2}S^{\rho\mu\nu}T_{\rho\mu\nu},
\ee
where $S^{\rho\mu\nu}=(K^{\mu\nu\rho}-g^{\rho\nu}T^{\sigma\mu}_{~~~\sigma}+g^{\rho\mu}T^{\sigma\nu}_{~~~\sigma})$,
 $h=det(h^a_{~\mu})$ and $k=8\pi G/c^4$. The Lagrangian is invariant under a local translation. The variation with respect to $B^a_{~\mu}$
 yields the field equation
 \be\label{fieldeq}
 \partial_{\sigma}(hS_a^{~\rho\sigma})-k(hj_a^{\rho})=0,
 \ee
 where
 \be
 hj_{a}^{\rho}=-\tfrac{\partial \mathcal{L}}{\partial B^a_{\rho}}
 \ee
 is the energy-momentum current.
 The field equation (\ref{fieldeq}) is equivalent to the Einstein field equations of GR.

Kleinert also suggests \cite{Kleinert2010} that GR (non-zero curvature; zero-torsion) and
teleparallel gravity (zero-curvature; non-zero torsion) are just the two extremes of a full class of
equivalent theories, all of which have non-zero curvature and non-zero torsion.

\section{The gravity Lie algebroid}

Teleparallel gravity provides a gauge theory for gravity which is analogous to a Yang-Mills theory.
The gauge group for Yang-Mills is $SU(n)$; for teleparallel gravity, as indicated in \cite{pereira2010} the group is the abelian Lie group of local translations
with the generators $\partial_a$. The main goal of this paper is to set teleparallel gravity in the framework of a fiber bundle.
The result we find is a generalized version of (the generators $\partial_a$ of) teleparallel gravity and
the corresponding geometrical structure is given by a Lie algebroid.

\subsection{The gauge generators $D_a$ and their algebra}

We start by considering (\ref{pocov}) $\nabla_a=h_a^{~\mu}\nabla_{\mu}$ and since we are interested in the translation part only,
we ignore the spin connection term, and rename it as follows
\be
D_a= h_a^{~\mu}\partial_{\mu}\, .
\ee
In this way it can be used to express the covariant derivative with respect to the potential for translations
\be
D_a=\partial_a+B_a
\ee
where $\partial_a=\delta_a^{~\mu}\partial_{\mu}$ and $B_a=B_a^{~\mu}\partial_{\mu}$.
We also notice that $h_a^{~\mu}\partial_{\mu}$ defines the non-holonomic basis $h_a$ of $TM$
\be\label{theha}
h_a=h_a^{~\mu}\partial_{\mu},
\ee
and conclude that
\be
D_a=h_a \, .
\ee
In this way the $D_a$ acquires the geometrical meaning of a basis, the dual of the $\theta^a$ found in \cite{Gronwald:1995em}.
At this point our work begins to differ from teleparallel gravity as found in \cite{pereira2010} since we assume the $D_a$ to be
 the \emph{generalized} generators of translations.
Direct computation reveals that
\be\label{dadb}
[D_a,D_b]= -T^{c}_{~ab}D_{c},
\ee
with
\be\label{Tcab}
T^c_{~ab}\equiv h_a^{~\mu}h_b^{~\nu}T^c_{~\mu\nu}.
\ee
Here $h_a^{~\mu}$ is used to convert between spacetime and internal indices \cite {Kibble1961,Blago2002}, and
the torsion field strength $T_{ab}\equiv T^c_{~ab}D_c$ assumes values in the algebra of the generators $D_a$.
Eq.(\ref{dadb}) gives \footnote{Expressed in components, the right-hand side reads:
 $[\partial_a B_b^{~\mu}-\partial_b B_a^{~\mu}+(B_a^{~\nu}\partial_{\nu}B_b^{~\mu}-B_b^{~\nu}\partial_{\nu}B_a^{~\mu})]h^c_{~\mu}D_c$}
\be
[D_a,D_b]=\partial_a B_b -\partial_b B_a+[B_a,B_b] \, .
\ee

 The commutator $[B_a,B_b]$ is non-zero, and it encapsulates the self interaction property of gravity.
The commutator between two tetrad bases is
\be\label{lieh}
[h_a,h_b]=f_{~ab}^{c}h_c,
\ee
in which the $f^c_{~ab}$ are the coefficients of anholomicity given by
\be
f_{~ab}^{c}=A_{~ba}^{c}-A_{~ab}^{c}-T_{~ab}^{c}.
\ee
In teleparallel gravity, the only contribution due to gravity to the connection $\Gamma ^{\rho}_{~\mu\nu}$
 comes from the Weitzenb\"{o}ck connection,
while the spin connection $A^a_{~b\mu}$ is zero.
We can make the identification (up to a sign)
\be
T_{~ab}^{c}(x)=f_{~ab}^{c}(x).
\ee
Because of (\ref{dadb}) and (\ref{lieh}) the torsion has two meanings: it is a field strength (gauge meaning),
and its components can be taken as structure `constants' (algebraic meaning).
This degeneracy arises since we treat the $a$'s as internal indices in order to describe the gauge aspect of the theory,
even though they are actually \emph{spacetime} and not internal \emph{color} indices as in Yang-Mills theory.
 We explicitly include the position dependance on $x\in M$ of both the torsion and the coefficients of anholomicity.
 This is crucial since now
we see that the components of the torsion can be interpreted as the structure functions of the algebra given by the $D_a$.
The structure `constants' become the structure functions depending on $x$.
The meaning of bases $D_a$ and their commutation relation (\ref{dadb}) has also been linked to noncommutative geometry as shown in \cite{Langmann:2001yr}.

\subsection{Lie algebroid}
Formally a Lie algebroid $E$ over a base manifold $M$ is the vector bundle $\pi:E\rightarrow M$ together with a map
$\rho:E\rightarrow TM$ called the anchor map of $E$ and an $\mathbb{R}$-linear Lie bracket $[\cdot,\cdot]_E$ defined on the space of sections $\Gamma (E)$
such that \cite{Weinb}
\be\label{defalgb}
\rho([u,v])=[\rho(u),\rho(v)]
\ee
\be\label{leib}
\qquad\qquad \left[u,fv\right]=f\left[u,v\right]+(\rho(u) f)v
\ee
for all $u,v\in \Gamma (E)$ and $f\in C^{\infty}(M)$.
A Lie algebroid is sometimes called an alternative tangent space,
    and this is due to the presence of the anchor map $\rho$ which induces a Lie algebra homomorphism (denoted with the same symbol $\rho$) between the
    Lie algebras of $\Gamma(E)$ and $\chi(M)$, $\rho: \Gamma(E)\rightarrow \chi(M)$, where $\chi(M)$ indicates
    the sections of $TM$.
In this way the anchor map makes $E$ behave very much like the tangent bundle: $\rho(u) = X$ with $X\in \chi(M)$.
 The bracket $[\cdot,\cdot]_E$ is not $C^{\infty}(M)$-linear due to the the presence of $\rho$, and therefore there is not a well
defined Lie algebra at each point $x\in M$. In the case of $\rho=0$ the Lie algebroid reduces to a Lie algebra bundle.

\subsection{The gravity Lie algebroid}

Consider a coordinate system $\{x^\mu\}$ on a chart in a manifold $M$.
 Let $\{D_a\}$ be a basis of sections in this chart of the Lie algebroid $(E,\rho,[\cdot,\cdot]_E)$.
   In this chart we can characterize the anchor map and Lie bracket by structure functions as

\be\label{str_fns} 
    [D_a(x),D_b(x)]_E =-T^c_{~ab}(x) D_c(x) \, .
    \ee
In the gravitational case the Lie algebroid is the tangent bundle, specifically:
\begin{itemize}
  \item the anchor map $\rho$ is the identity map such that $\rho(D_a) = h_a $.
  \item the elements $u,v\in \Gamma(E)$ are mapped to the vector fields $X,Y\in \chi(M)$ via the anchor map $\rho$.
  \item the fibers (spanned by the basis $D_a$) are copies of the tangent space (spanned by the basis $h_a$).
  \item the term ($\rho(u)f$) in (\ref{leib}) corresponds to $\mathcal{L}_X f$, the Lie derivative of $f$ with respect to $X$.
\end{itemize}

We call this Lie algebroid the Lie gravity algebroid. The geometrical structure used to study the gauge aspects of gravity is not
 a principle bundle, in contrast to the case for Yang-Mills theories, but it is simply the tangent bundle seen as a Lie algebroid
 in which the gauge generators $D_a$ are taken to be the bases of the $\Gamma(E)$.
 It is useful to express the commutations between two elements of the algebra in components, by the use of (\ref{leib})
 \be\label{ldnonc}
 [X^a h_a,Y^b h_b]=X^aY^b[h_a,h_b]+ (X^a h_a(Y^b)-Y^a h_a(X^b))h_b,
 \ee
 which is $\mathcal{L}_X Y$ expressed in a non-coordinate basis.
 In the presence of gravity, the torsion enters directly in the first term on the RHS of (\ref{ldnonc}) because of (\ref{str_fns}).
 The quantity $\varepsilon^{a}(x^{\mu})D_a$ is an element of the translation algebra with $\varepsilon^a(x^{\mu})$
 the infinitesimal parameter for local translation (see Eq. (\ref{ltra})).

If there were no gravity\footnote{A Universe without gravity is possibly an unrealistic nonphysical situation since everything which exists has energy
  and therefore is a source of gravity.},
  the torsion $T^c_{~ab}(x)$ and the gauge potential $B_a(x)$ are zero, the generators, or bases of the sections,
 are given by the $\partial_a$ and the $u,v$ $\in \Gamma(E)$ have components $\varepsilon^a$ which are constants, i.e global translations.
 This is the well defined abelian Lie algebra of global translations.

\subsection{The gravity Lie groupoid}
A groupoid generalizes the concept of a group in the sense that the multiplication of two distinct groupoid elements exists only
if certain conditions are met.
Specifically a groupoid consists of the two sets
 \footnote{Some authors use the term \emph{groupoid} to indicate the set $\Gamma$ only, others use it to indicate the whole structure ($\Gamma,B$ and $s,t$)}:
 the \emph{groupoid}  $\Gamma$ and the \emph{base} $B$,  together with a pair of \emph{maps} $s,t:\Gamma \rightarrow B$ called respectively the \emph{source }and the \emph{target} map.
The $g_1,g_2, ...$ are the groupoid elements of $\Gamma$, and $b_1,b_2 ,...$ are the base elements of $B$, such that $s(g_1)=b_1$ and $t(g_1)=b_2$.
In this way $g_1$ can be seen as a map acting on the $b\in B, ~g_1:b_1 \rightarrow b_2$.
Given $g_1$, and $g_2$ the product between the two exist iff $t(g_1)=s(g_2)$
\footnote{For the full definition which includes the \emph{inverse} map, the \emph{inclusion} map, the \emph{units} and theirs proprieties see
\cite{Weinb,Mayer:1989ng}}.
In the case of a Lie groupoid both $\Gamma$ and $B$ are differential manifolds and $s,t$ are smooth functions (surjective submersions) which define the
fibers $\Gamma_{b_1}=s^{-1}(b_1), \Gamma^{b_1}=t^{-1}(b_1)$. 

Similarly to the relation between Lie groups and Lie algebras, a Lie algebroid contains the local information of a Lie groupoid.
In the case of the gravity Lie algebroid, the corresponding groupoid can be taken as the pair groupoid $\Gamma = M\times M$ or the fundamental
groupoid $\Pi(M)$ which locally is isomorphic to $M\times M$.
The pair groupoid has elements given by the pair (x,y).
 The fundamental groupoid of a manifold is a generalization of the fundamental group in which the elements are given by
 the equivalence classes of paths between two points instead of loops at one point.
 Formally an element $g_{xy}\in \Pi(M)$ can be written as $(x,[\gamma],y)$
 with $x,y\in M$, $\gamma$ represents a path between $x$ and $y$, and $[\gamma]$ is the equivalence class of these paths, such that:
 \be
 s(g_{xy})=x\qquad t(g_{xy})=y.
 \ee
 In this way the composition between the two elements $g_{xy}$ and $g_{yz}$ corresponds to the concatenation of paths which exists
  since $t(g_{xy})=s(g_{yz})$ and it gives $g_{xz}$.
In the case s([$\gamma$]) = t([$\gamma$]) = $x$ and s($[\delta]$) = t($[\delta]$) = $y$ for each path $\gamma, \delta$ ... and
$x, y, ... \in M$, the paths become loops, and the fundamental groupoid reduces
to the fundamental group of $M$ at $x$.

A formal definition of the exponential map for a Lie groupoid can be found in \cite{mackenzie05} where the
pair groupoid is given as an example.

\subsection{Observables in the theory of gravity}

The gravity Lie algebroid gives new perspective to the meaning of an observable quantity.
In established gauge theories such as electromagnetism and Yang-Mills theory,
 observable physical quantities are gauge-invariant quantities, for example
the field strength of electromagnetism $F_{\mu\nu}$, which is a local $U(1)$ gauge invariant quantity.
The gravitational case is fundamentally different.
 For a general gravitational field,
the structure `constants' $T(x)_{~ab}^{c}$ depend on the position and the anchor map $\rho$ is not zero,
i.e. there is not a well defined Lie algebra at the points in $M^4$. This implies that there is not
a gauge group, and therefore there are no symmetries with respect to which a quantity can possibly be gauge-invariant.
We interpret this to be the reason why the theory of gravity does not possess intrinsic observable quantities in the general case.
By intrinsic observable quantities of the gravitational field we refer to quantities such as the energy density,
the momentum (examples: a black hole moving in space, gravitational waves),
and angular momentum (of a black hole for example).
They can all be derived from the components of the Riemann tensor (in GR).
 These quantities (or a generalization of their meaning)
could become observable if a generalized version of Noether's theorem for groupoids exists.
Since we have found that the geometrical structure for gravity is a Lie algebroid,
the concept of symmetry seems to fail: there is a different symmetry at each point
which is equivalent to no symmetry at all. Our case is fundamentally different
from the localization of the gauge parameter as for example for
electromagnetism. In the latter case when the gauge parameter $\theta$ becomes $\theta(x)$ the theory is still gauge invariant
with respect to $U(1)$.
If there are no symmetries, according to Noether's theorem, there should be no conserved currents and charges.
This issue, in the case of gravity, seems to be strictly connected with the problem of defining the energy density of the gravitational field.
In order to develop dynamics, a different and more general concept of symmetry (and gauge invariance) for groupoids seems necessary.
This would also require a generalized version of Noether's theorem to define and give an interpretation to the conserved quantities.

Nevertheless it is possible for these quantities to be observable when a group (i.e. a symmetry) is present.
This happens in two cases.
The first case is when the group of symmetries is within the gravitational field and specifically
when the asymptotic behavior of the gravitational field possesses symmetries
(for example Minkowski, AdS). It is well known that the local energy density of the gravitational field
is not well-defined in general, but the total energy over a volume can be defined
for asymptotic Minkowski spacetimes (the ADM mass) \cite{Wald:1984rg}.

The second case is directly related to the hole argument: giving a physical meaning to a spacetime point is possible
when gravity is coupled to electromagnetism or Yang-Mills, i.e. if there is also a non-gravitational interaction, as for example a collision \cite{Rovellib}.
The symmetry group in this case is $SU(n)$ and it becomes possible to define and measure the numerical value of a gravitational observable,
 which allows the specification of \emph{where} for example two photons interact.

%

\section{Conclusions}
In this paper we present the Lie algebroid (and the corresponding Lie groupoid) for gravity as a new geometrical structure on which the theory of gravity is set.
To put our results in perspective, in the development of our theory we are at the same point  which existed at the beginning of the development of general relativity.
What we have found is that a manifold with a Lorentzian signature should be used, but
 we do not know that the presence of gravitation should be represented by a metric, and we have yet to obtain the equations which describe dynamics.
In other words, in this paper we present the geometrical background on which gravity is set and how its gauge aspects arise.
One of the main issues related to understanding dynamics in the context of algebroids, is the the validity of Noether's theorem.
Of course we are conducting further research in this direction.

Our results also provide a basis for various directions of research.
An interesting issue is to investigate the relations between the gravity Lie algebroid (of local translations with different generators at each point) and
the ordinary Yang-Mills algebras based on $SU(N)$ for the physical situations where both interaction are present.
This leads to the theory of algebra deformation since the components of the torsion depend on the point $P$ at which the torsion is measured.
It's also interesting to notice how the $D_a$ correspond to the \emph{displacement algebra} generators $d_l$ proposed in \cite{Chan1993}
in which the $d_l$ generalize the translation algebra to the case of a non abelian base manifold and
the corresponding structure constants are also given by the torsion.
Their model aims to extend Yang-Mills theory with such a base manifold and the expressions of the gauge potential and field strength are shown.
However, gravity is not explicitly included in \cite{Chan1993} (the torsion does not acquires the gravitational meaning as found in teleparallel gravity)
and the development of our model (\ref{dadb}) might provide the coupling of Yang-Mills theory and gravity in this scenario.
Alternative constructions to a principal bundle, especially for gravity and its gauge properties have already been considered, such as for the case of
composite fiber bundles \cite{Tresguerres2002}. Correspondences and equivalences between the gravity algebroid and the composite fiber bundle seems a
natural and interesting line of investigation.
The use of a groupoid to describe Yang-Mills-like gauge theory has been proposed in \cite{Mayer:1989ng,Guay:2004zz}, especially to investigate the quantum aspects
of the theory. In this case there exists a groupoid associated to the principal bundle named the \emph{gauge} groupoid. This suggests another possible
line of research: the construction of a similar gauge groupoid for the gravity Lie groupoid (the fundamental groupoid).

The main goal is a quantum theory of gravity. Given the richness of the Lie algebroid construction \cite{Weinb,Saemann:2012ab} and specifically the connections
with non commutative geometry \cite{Langmann:2001yr}, we claim that the results of this paper
 represent a first step in this direction.

\section{Acknowledgments}
S. F. would like to point out the following: the main results and
intuitions present in this paper, as for example the equation
(\ref{str_fns}), originated during collaborations with his dear
friend George Stephen Karatheodoris who passed away suddenly on
September 22, 2012 at the age of 37. They represent only a part of
the brilliant ideas he had toward a quantum description of gravity.
It has been a privilege for S.F. to work with him for the past five
years and, to honor his research, to present and develop this
project. S.F. would like to thank A. Stern for useful
and clarifying discussions on several topics.

We also would like to thank A. Weinstein for his helpful considerations and suggestions.
The work of B.H. and S.H. is supported in part by the DOE under grant DE-FG02-10ER41714.




\end{document}